\documentstyle[12pt,epsfig]{article}
\begin{document}
\begin{center}
{\Large{\bf{ATLAS Sensitivity to the Flavour-Changing
Neutral Current Decay $t \rightarrow Zq$}}}
\end{center}
\   \par
\   \par
\begin{center}
{\large{\bf{Leila Chikovani}}}
\end{center}
\begin{center}
Institute of Physics of the Georgian Academy of Sciencies,\par
\end{center}
\   \par
\   \par
\begin{center}
{\large{\bf{Tamar Djobava}}}
\end{center}
\begin{center}
High Energy Physics Institute, Tbilisi State University,\par
\end{center}
\  \par
\  \par
\begin{center}
{\bf{ABSTRACT}}
\end{center}
\  \par
\  \par
The sensitivity of the ATLAS experiment to the
top-quark rare decay via flavor-changing neutral currents
$t \rightarrow Zq$ ($q$ represents $c$ and $u$ quarks) has been studied
at $\sqrt{s}$=14 TeV in two decay modes:

1.The pure leptonic decay of gauge bosons:
$t\bar{t} \rightarrow ZqWb \rightarrow l^{+} l^{-} j l^{\pm} \nu b$ ,
(l=e, $\mu$).

2.The leptonic decay of Z bosons and hadronic decay of W bosons:
$t\bar{t} \rightarrow ZqWb \rightarrow l^{+} l^{-} jjjb$ ,
(l=e, $\mu$). 

The dominant backgrounds $Z+jets$, $WZ$ and $t\bar{t}$ have been analysed. 
The signal and backgrounds were generated via PYTHIA 5.7, and simulated  
and analysed using ATLFAST 2.14. A branching ratio for $t \rightarrow Zq$ 
as low as 2.0x10$^{-4}$ for the leptonic mode and  5.9x10$^{-4}$ for 
hadronic mode could be discovered at the 5$\sigma$ level with an integrated 
luminosity of 100 fb$^{-1}$.

\newpage

\section{Introduction}

The existence of the top quark has been established at the Fermilab
Tevatron by the CDF and DO Collaborations [1]. Due to the large value of 
the top quark mass, of order the Fermi scale, the top quark couples 
strongly to the electroweak symmetry breaking sector. If anomalous top 
quark couplings beyond the Standard Model (SM) exist, they could affect top quark 
production and decay processes at hadron and $e^{+}$ $e^{-}$ colliders [2,3].
For example, study of the flavour-changing neutral current top quark
decay $t \rightarrow Zq$ (where $q$ represents either $c$ or $u$ quarks)
is of great interest.  As is well known [4,5], the LHC can be considered as
a ``top factory'', producing about 80,000
$t\bar{t}$ events per day at $L=10^{33}~cm^{-2}~s^{-1}$, 
making the LHC an ideal place to explore this rare decay.

In this note, we present a study of 
the sensitivity of the ATLAS experiment to the branching ratio
of the top-quark rare decay mode $t \rightarrow Zq$ ($q=u,c$).
In the framework of the Standard Model,
the loop suppression and heaviness of gauge bosons make this process
extremely rare.  The SM prediction for the branching ratio, of order 10$^{-13}$ [6],
would render the decay unobservable even with the large samples of top
quarks expected at the LHC.  However, other models predict
significantly larger branching ratios for this process. 
For example, in two-Higgs-doublet
models, the value of Br($t \rightarrow Zq$) can reach $\sim 10^{-9}$ [7],
while Supersymmetric (SUSY) models (without $R$-parity) can have even much
higher values, Br($t \rightarrow Zq$)$\sim 10^{-4}$ [8]. 

Experimentally, little is known about this decay.  The only
existing limit comes from a CDF analysis of its Run1 data [9],
yielding Br($t \rightarrow Zq$) $< 33\%$ (95$\%$ CL).  Given the
SM prediction, an observation of this decay mode, even at the LHC,
would provide a clear signal of new physics beyond the SM, such as new 
dynamical interactions of top quark,
multi-Higgs doublets, exotic fermions or other possibilities [5,6,7].
In addition, this mode is of interest since, if the branching ratio were
large, events of the form
$t\bar{t} \rightarrow ZZ + cc$ could prove to be a serious background
to events containing $Z$ boson pairs and jets from cascade decays of
squarks and gluinos [10].

The dominant mechanism for top quark production at the LHC is
$t \bar t$ pair production via $gg, q\bar{q} \rightarrow t\bar{t}$.
The analyses presented here focus on two different  
final state topologies of $t\bar{t} \rightarrow ZqWb$: \\
1. The purely leptonic decay of both gauge bosons:\\
$t \rightarrow Zq \rightarrow l^{+} l^{-} j$ ,
$\bar{t} \rightarrow Wb \rightarrow l^{\pm} \nu b$ (l=e, $\mu$)\\
2. The leptonic decay of the Z boson and hadronic decay of the W boson:\\
$t \rightarrow Zq \rightarrow l^{+} l^{-} j$ ,
$\bar{t} \rightarrow Wb \rightarrow jjb$ (l=e, $\mu$)\\

The value of the branching ratio of the second (``hadronic'') mode is three 
times higher than for the first (``leptonic'') one.  However, the hadronic mode
suffers from enormous QCD-backgrounds, while the leptonic mode has a very 
distinct experimental signature.  Both modes have been studied, and the results 
of each analysis will be presented below.

\section{Monte Carlo Event Generation}

The dominant source of top quark production at the LHC is pair-production via
$gg, q\bar{q} \rightarrow t\bar{t}$.
PYTHIA 5.7 was set up to produce $t\bar{t}$ events
at $\sqrt{s}=14$ TeV and $m_{top}=174$ GeV, with proton structure
functions CTEQ2L. Initial and final state QED and QCD (ISR, FSR) radiation,
multiple interactions, fragmentations and decays of unstabled particles
were enabled.

The decay $t \rightarrow Zq$ is not implemented in the
standard release of PYTHIA. To include this process in PYTHIA,
all individual decay channels of the top quark were first switched off,
except for $t \rightarrow Wb$ and $t \rightarrow Ws$. The channel 
$t \rightarrow Ws$ was then replaced by the decay $t \rightarrow Zq$, 
by replacing  $W$ by $Z$ and $s$ by $c(u)$.  The total cross-section for
$t \bar t$ production was assumed to be $\sigma_{t\bar{t}}$=800 pb [12].

The analyses reconstructed the $Z$ boson via its leptonic decay,
$Z \rightarrow l^+l^-$.  The experimental signature included, therefore, 
a pair of isolated charged leptons, as well as several jets.  The
SM backgrounds considered for this process were $Z+jets$ production, $WZ$ 
production, and $t \bar t \rightarrow WbWb$ production.

$Z+jets$ production at the LHC has a relatively large 
cross-section, dominated by $qg \rightarrow Zq $ and $q\bar{q} \rightarrow 
Zg$ processes. To decrease the size of the background sample which needed
to be generated, thresholds were imposed at the generator level on 
the invariant mass, $\hat{m}=\sqrt{\hat{s}} > 130$ GeV, and 
transverse momentum, $\hat{p_{\perp}} > 50$ GeV, of the hard
scattering process. The cross-section for this subsample of events 
was $\sigma_{Z+jets}$=3186 pb. 

The $WZ$ background  is the electroweak process $pp \rightarrow W^{\pm}Z+X$,
and has an assumed cross-section of $\sigma_{WZ}$ = 26.58 pb.

The performance of the ATLAS detector was simulated using the
fast simulation package ATLFAST 2.14 [11], which uses parametrizations
of the detector resolution functions. 

\section{The Leptonic Decay Mode}

The final state for the leptonic decay mode is
$t\bar{t} \rightarrow ZqWb \rightarrow l^+l^-j l\nu b$.
The experimental signature therefore includes three isolated charged 
leptons, two of which reconstruct a
$Z$ boson, and large missing transverse energy due to the neutrino.

The backgrounds considered were $Z + jets$ production, followed
by the decay $Z \rightarrow l^+l^-$,
$pp \rightarrow W^{\pm}Z+X \rightarrow l^{\pm}\nu l^+l^- +X$,
and $t\bar{t} \rightarrow W^{+}bW^{-} \tilde {b}
\rightarrow l^{+} \nu b l^{-} \tilde {\nu} b$.
Background samples of $2.1 \times 10^7 Z + jets$ events, 
35700 $WZ$ events, and $3.7 \times 10^6$ $t \bar t$ events
were generated.
Assuming the production cross-sections given earlier, and 
including the relevant branching ratios, these background
samples correspond to an integrated luminosity of 100 fb$^{-1}$.

Table \ref{tab:leptonic} summarizes the effects of the sequential 
application of the various analysis
cuts on the background samples and on the sample of 20565 signal
events of the topology 
$t\bar{t} \rightarrow ZqWb \rightarrow l^+l^-j l\nu b$.

\begin{table}[h]
\begin {tabular}{|l||c|c||c|c||c|c||c|c|}\hline
\multicolumn{1}{|c||}{Description}&\multicolumn{2}{|c||}
{$t \rightarrow Zc$}&\multicolumn{6}{|c|}{Background Processes}\\ \cline{4-9}
\multicolumn{1}{|c||}{of}&\multicolumn{2}{|c||}
{Signal}&\multicolumn{2}{|c||}{Z+jets}&
\multicolumn{2}{|c||}{Z+W}&\multicolumn{2}{|c|}{$t \bar t$}\\ \cline{2-9}
\multicolumn{1}{|c||}{Cuts} & Nevt & Eff (\%) & Nevt & Eff (\%) 
& Nevt & Eff (\%) & Nevt & Eff (\%) \\ \hline \hline
Nevt gen. &20565& &2.1$\cdot 10^{7}$& &35000& &3.7$\cdot 10^{6}$&  \\ \hline
Preselection & 16497 & 80.2 & $3.7\cdot 10^{5}$ & 1.7 & 2941 & 8.2 
& 8.5$\cdot 10^{5}$ & 29.4 \\ \hline
3 leptons & 8885 & 43.3 & 945 & 4.4$\cdot10^{-3}$ & 1778 & 5.0 & 1858 & 
5.0$\cdot10^{-2}$ \\ \hline 
$\not\!P_{T}>30~GeV$ & 6730 & 32.7 & 80 & 3.7$\cdot10^{-4}$ & 1252 & 3.5 
& 1600 & 4.3$\cdot10^{-2}$ \\ \hline 
2 jets & 4063 & 19.8 & 31 & 1.5$\cdot10^{-4}$ & 225 
& 0.6 & 596 & 1.6$\cdot10^{-2}$ \\ \hline 
$M_{Z} \pm 6~GeV$ & 3450 & 16.8 & 24 & 1.1$\cdot10^{-4}$ & 180 
& 0.5 & 29 & 7.8$\cdot10^{-4}$ \\ \hline
one $b$-tag & 1678 & 8.2 & 10 & 4.7$\cdot10^{-5}$ & 28 
& 0.04 & 10 & 2.7$\cdot10^{-4}$ \\ \hline \hline 
$M_t \pm 8~GeV$& 728 & 3.5 & 0 & 0 & 1 
& 2.8$\cdot10^{-3}$ & 3 & 8.1$\cdot10^{-5}$ \\ \hline
$M_t \pm 12~GeV$& 973 & 4.7 & 0 & 0 & 1 
& 2.8$\cdot10^{-3}$ & 4 & 1.1$\cdot10^{-4}$ \\ \hline
$M_t \pm 24~GeV$& 1264 & 6.1 & 0 & 0 & 2 
& 5.6$\cdot10^{-3}$ & 5 & 1.3$\cdot10^{-4}$ \\ \hline
\end{tabular}
\caption{The numbers of events and efficiencies ($\%$) of kinematic cuts
applied in sequence for the signal and backgrounds for the leptonic mode.  
The numbers do not include the lepton identification efficiencies.}
\label{tab:leptonic}
\end{table}

Preselection cuts were first applied, requiring the presence of at 
least three charged leptons (electrons with $p_{T} > 5 $ GeV and muons with
$p_{T} > 6 $ GeV) within pseudorapidity $|\eta| < 2.5$.  Of
these, at least one pair of leptons must be of opposite sign and same flavour,
compatible with them being produced from a $Z$ decay.  In addition, 
the number of jets in the event with $p_{T}^{jet} > 15 $ GeV 
was required to be at least two. The requirement of three leptons reduces
significantly the $Z+jets$ and $t\bar{t}$ backgrounds, while the requirement
of two jets reduces significantly $WZ$ and $Z+jets$ backgrounds.

\begin{figure}[hp]
\begin{center}
\epsfig{file=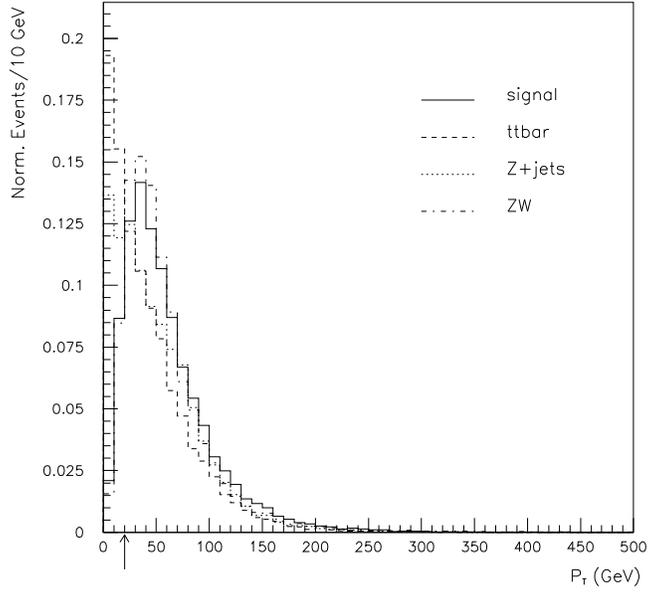,width=8.5cm}
\end{center}
\vspace{-0.5cm}
\caption{The $p_{T}$ distributions of leptons which give a
dilepton invariant mass within the accepted window around the $Z$ mass,
for leptonic mode signal and backgrounds, normalised to unity.
The arrow indicates the threshold value $p_{T}^{l}$ used for 
the analysis cut.}
\end{figure}

\begin{figure}[hp]
\begin{center}
\epsfig{file=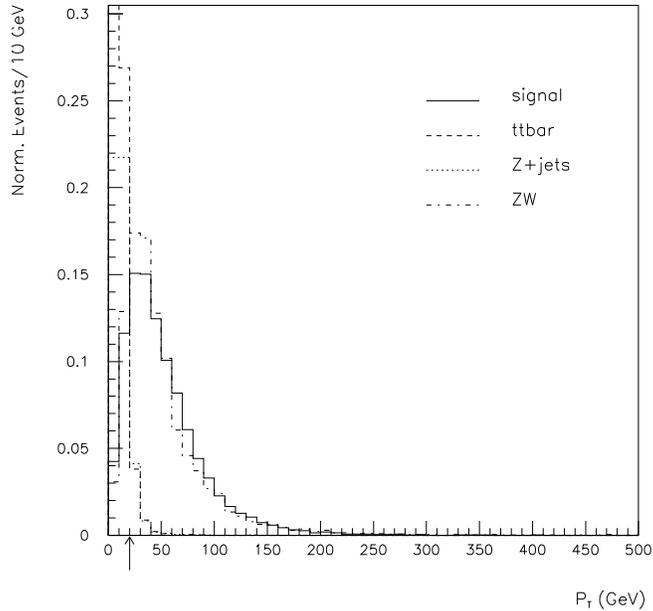,width=8.5cm}
\end{center}
\vspace{-0.5cm}
\caption{The $p_{T}$ distribution of the third lepton for the leptonic
mode signal and backgrounds, normalised to unity. The arrow indicates 
the threshold value $p_{T}^{l}$ used for the analysis cut.}
\end{figure}

The lepton criteria were then tightened, by requiring the presence of
at least three isolated, charged leptons (electrons or
muons) with $p_{T}^{l} > 20$ GeV and $|\eta^{l}| < 2.5$. 
The isolation criterion required that there be no additional track with 
$p_{T} > 2$ GeV in a cone of ${\Delta}R=\sqrt{(\Delta\phi)^2+(\Delta\eta)^2}
= 0.3$ around the lepton direction. 
Fig. 1 presents the $p_{T}$ distribution of the two leptons
consistent with originating from a $Z$ decay for signal and backgrounds,
while Fig. 2 shows the $p_{T}$ distribution for the third lepton. 
Tightening the lepton criteria significantly reduces the 
$Z+jets$ and $t\bar{t}$ backgrounds, where any third lepton found 
typically comes from cascade decays of $b$-jets.

The next requirement, namely that the missing transverse momentum
in the event satisfies $p_{T}^{miss} > 30$ GeV, is effective
at further reducing the $Z + jets$ background (see Fig. 3),
while having little impact on the signal and other background sources.

\begin{figure}[h]
\begin{center}
\epsfig{file=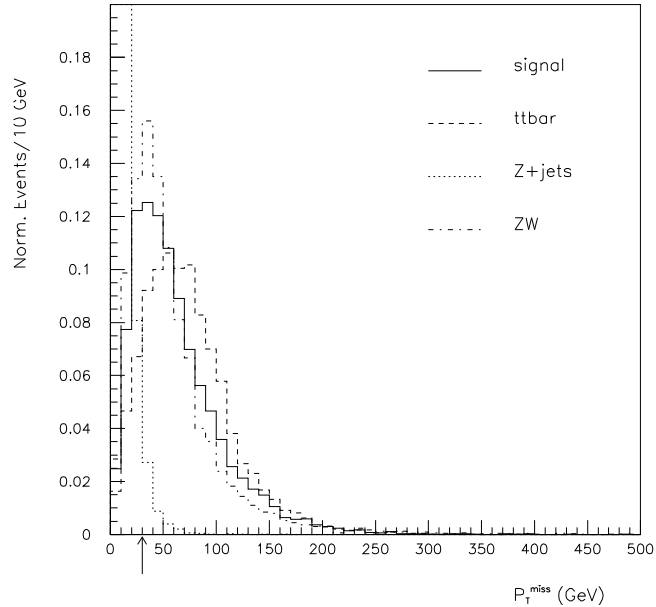,width=8.5cm}
\end{center}
\vspace{-0.5cm}
\caption{The reconstructed $p_{T}^{miss}$ distributions for leptonic
mode signal and backgrounds, normalised to unity.
The arrow indicates the threshold value used for the kinematic cut.}
\end{figure}

\begin{figure}[h]
\begin{center}
\epsfig{file=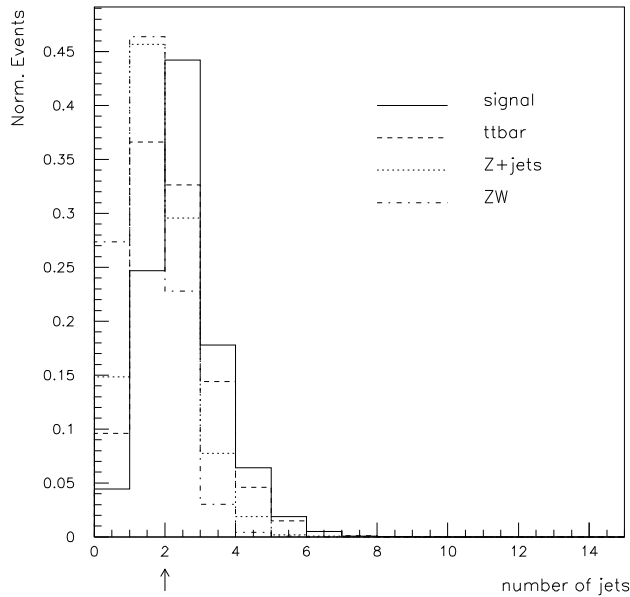,width=8.5cm}
\end{center}
\vspace{-0.5cm}
\caption
{Distribution of jet multiplicity (threshold at $p_{T}^{jet} > 50$ GeV)
for signal and backgrounds, normalised to unity.
The arrow indicates the threshold value of the number of jets
used for the kinematic cut.}
\end{figure}

Next, it was demanded that there be at least two jets with 
$p_{T}^{jet} > 50$ GeV,
$|\eta^{jet}| < 2.5$, and satisfying the following isolation conditions: 
${\Delta}R_{jj} > 0.4$ (jet-jet isolation) and ${\Delta}R_{lj} > 0.4$ 
(lepton-jet isolation). Fig. 4 presents the number
of jets with $p_{T}^{jet} > 50$ GeV in an event. One can see that 
the cut requiring the presence of two and more such jets in each event 
effectively suppresses the $WZ$ background.

The presence of a reconstructed $Z \rightarrow l^+l^-$ decay is
a powerful cut against the $t \bar t$ background.  A like-sign,
same-flavor pair of isolated leptons was required to reconstruct to
the $Z$ mass within a window of $M_{Z} \pm 6$ GeV mass window. 
Fig. 5(a) presents the distribution
of reconstructed invariant mass of $ll$ pairs $m_{ll}$, for 
all dilepton combinations for the signal events.  The width of the accepted 
window corresponds to approximately twice the $Z$ mass resolution
of about 2.9 GeV. 

The next requirement was the presence in the event of 
one tagged  $b$-jet, which is effective at further reducing
the $WZ$ background (see Fig. 6).

Finally, a peak at the top quark mass in the $Zj$ invariant mass
distribution was sought. In Fig. 5(b), the distribution
of reconstructed invariant mass $m_{llj}$ for all combinations of $llj$
is presented for the signal events. The top quark mass 
resolution is $\sigma(m_{llj}) = 14$ GeV.  Accepted combinations
were required to lie within a window around the known top quark
mass.  As a check of the stability of the results, three
different mass window were considered:  
$m_{Zj}$ = $m_t \pm 8$ GeV (narrow cut), $m_t \pm 12$ GeV 
($\sim ~ \sigma$), and $m_t$ $\pm 24$ GeV ($\sim ~2\sigma$).
The top mass window removes almost completely the remaining
background.  

\begin{figure}[hbpt]
\begin{center}
\epsfig{file=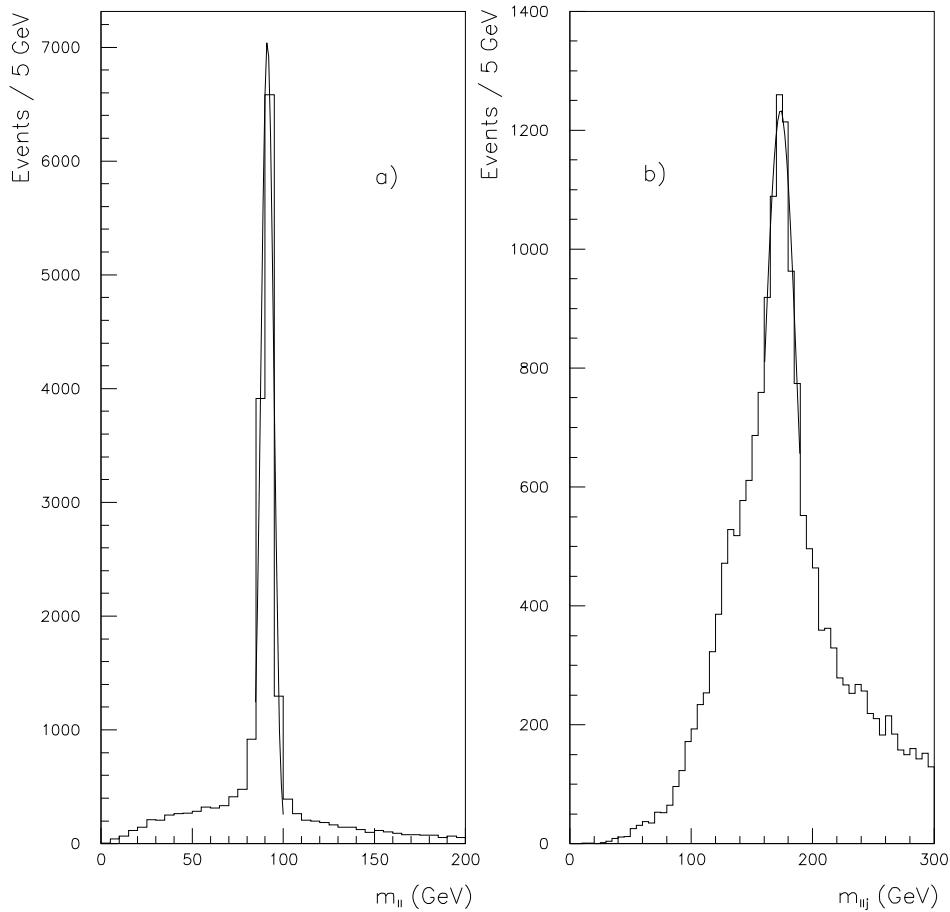,width=14cm}
\end{center}
\vspace{-0.5cm}
\caption
{(a) Distribution of reconstructed invariant mass of the lepton pairs,
$m_{ll}$ for the leptonic mode.
(b) Distribution of reconstructed invariant mass of
$t \rightarrow Zq \rightarrow llj$ for the leptonic mode.}
\end{figure}

\begin{figure}[hbpt]
\begin{center}
\epsfig{file=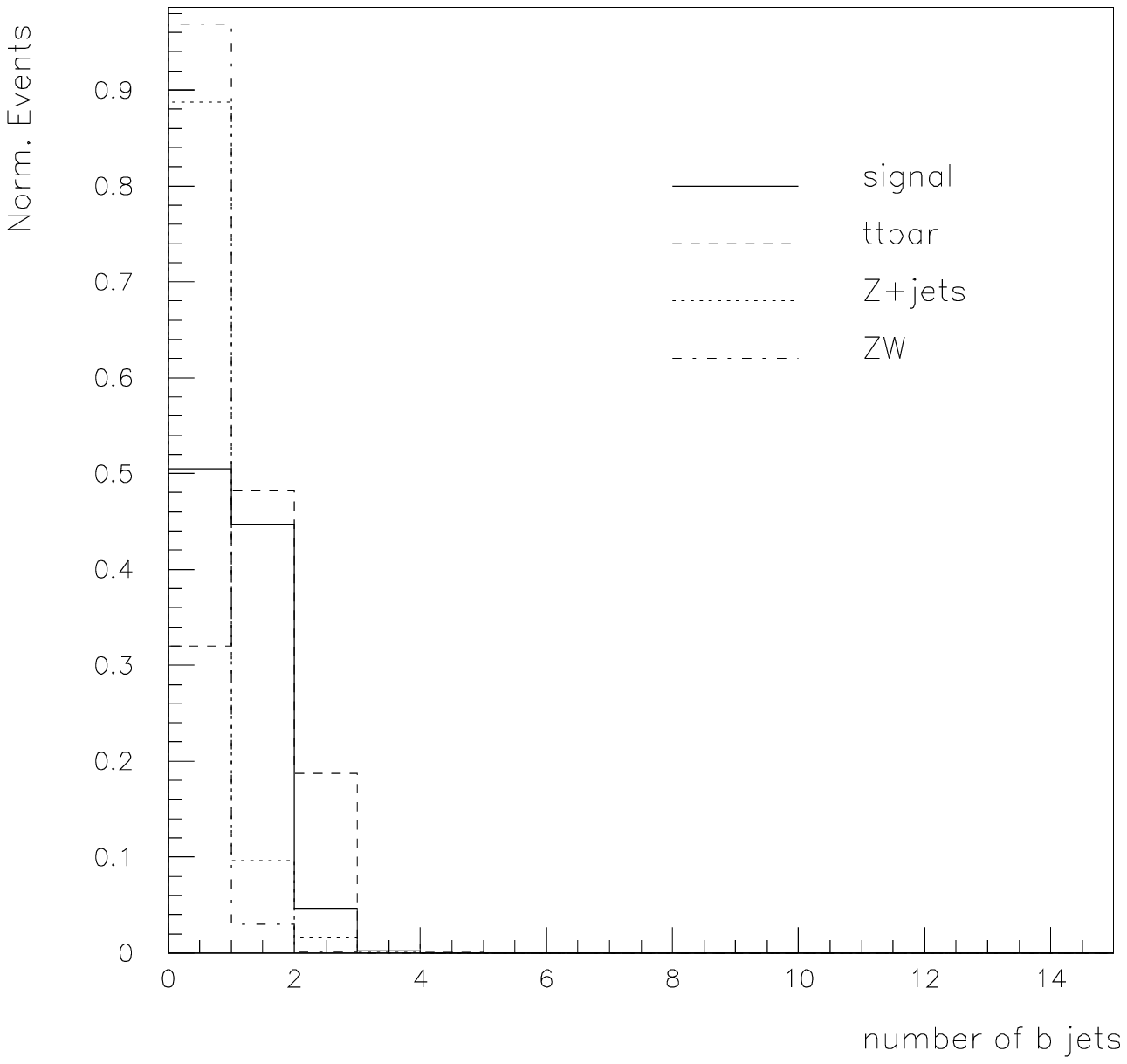,width=9cm}
\end{center}
\vspace{-0.5cm}
\caption
{Distribution of $b$-jet multiplicity (threshold at $P_{T}^{jet} > 50$ GeV)
for signal and backgrounds, normalised to unity.}
\end{figure}

As summarized in Table \ref{tab:leptonic}, the
final signal efficiency was 6.1\% for the $m_t$ $\pm 24$ GeV
mass window, with a total background of 7 events for
an integrated luminosity of 100 fb$^{-1}$.  The results
in the table do not yet include the lepton identification
efficiencies.  Assuming an efficiency of 90\% for each of
the three leptons, the results correspond to a 5$\sigma$
discovery potential for 
Br($t \rightarrow Zq$) as low as 2.0x10$^{-4}$ and for
an integrated luminosity of 100 fb$^{-1}$.  This result is
in good agreement with that of reference [13].

\section{The Hadronic Decay Mode}

The final state for the hadronic W decay mode is 
$t\bar{t} \rightarrow Zc Wb \rightarrow l^+l^-j jjb$.
The experimental signature
for the hadronic mode includes, therefore, two isolated charged 
leptons which originate from the $Z$ decay, and four energetic jets.

This mode has the following backgrounds:
$Z( \rightarrow ll) +jets$, $pp \rightarrow W^{\pm}Z+X 
\rightarrow jj l^+l^- +X$, and $t\bar{t} \rightarrow WbWb$ with the
final state topologies (a) $l^+\nu b l^- \tilde {\nu}b$, (b) $jjb jjb$,
or (c) $l^{\pm} \nu b jjb$.
In the case of (a),  the additional two jets must come from QCD radiation, 
while in (b) and (c) the source of leptons is from cascade decays. 
Therefore, requiring two isolated
energetic leptons and four energetic jets, among which only one 
is tagged as a $b$-jet, 
significantly suppresses the $t \bar t$ background. It has been
checked that the small remaining $t \bar t$ background is removed by the
reconstruction cuts asking for $m_{ll}$ and 
$m_{llj}$ to lie within windows around $m_Z$ and $m_t$ respectively.
Therefore, the $t \bar t$ background is reduced to a negligible
level, and is not considered further.

Background samples of $2.1 \times 10^7 Z + jets$ events, 
and $1.2 \times 10^5 WZ$ events were generated.
Assuming the production cross-sections given earlier, and 
including the relevant branching ratios, these background
samples correspond to an integrated luminosity of 100 fb$^{-1}$.

Table \ref{tab:hadronic} summarizes the effects of the sequential 
application of the various analysis
cuts on the background samples and on the sample of 19000 signal
events of the topology $t\bar{t} \rightarrow ZqWb \rightarrow l^+l^-j jjb$.

The analysis began with preselection cuts requiring that the
event contains at least two charged leptons (electrons with $p_{T} > 5 $ GeV
within pseudorapidity $|\eta| < 2.5$ and muons with
$P_{T} > 6 $ GeV within pseudorapidity $|\eta| < 2.4$), and include
a pair of opposite-sign and same-flavour
leptons, compatible with them having come from a $Z$ decay.
In addition, the number of jets with $p_{T}^{jet} > 15 $ GeV was
required to be at least four.
After preselection cuts, 46\% of the signal events are accepted,
while only 3.5\% and 4.1\% of the 
$Z+jets$ and $WZ$ background events, respectively, are retained.

The next cuts required the presence of two isolated leptons with
$p_{T}^{l} > 20$ GeV and $|\eta^{l}| < 2.5$, and the demand for
at least four jets with $P_{T}^{jet} > 50$ GeV and $|\eta^{j}| < 2.5$.
Jet-jet and lepton-jet isolation criteria were then applied,
as was done in the leptonic mode analysis.

A cut was then placed on the dilepton invariant mass, requiring that it
lie within a window of $\pm$6 GeV around $m_Z$, the same mass window as
used for the leptonic mode. 
Fig. 7(a) presents the distribution of reconstructed
dilepton invariant mass for the signal sample. 

To suppress the large remaining $Z+jets$ background, it was necessary
to use the information that signal events contain, in addition
to the decay $t \rightarrow Zq$, a hadronic decay $t \rightarrow
Wb \rightarrow jjb$ of the other top quark.  The hadronic top
quark decay was, therefore, reconstructed as part of the
signal requirement.  First, a pair of jets was required to
have an invariant mass $m_{jj}$ with a window of $\pm 16~GeV$ around $m_W$.  
Fig. 8.a shows the distribution of reconstructed
$m_{jj}$ for the best combinations of $jj$ for the
signal events. The $W$ mass resolution is $\sigma_{m_{jj}}$ = 8 GeV. 

The requirement was then made to have one jet tagged as a $b$-jet.
Finally, the $jjb$ invariant mass was required to lie within a
window of $\pm 8$ GeV around $m_t$.
Fig. 8(b) presents the distribution of the reconstructed invariant top
mass ($m_{jjb})$ for the best combinations of $jjb$
for the signal. The top mass resolution is $\sigma (m_{jjb})$ = 18.5 GeV,
implying that the mass window applied is rather narrow in order to
get a large background rejection.
The sequence of cuts required to reconstruct the hadronic decay
of the other top quark dramatically suppresses the background,
but also reduces the signal efficiency by almost an order
of magnitude.

\begin{figure}[hbpt]
\begin{center}
\epsfig{file=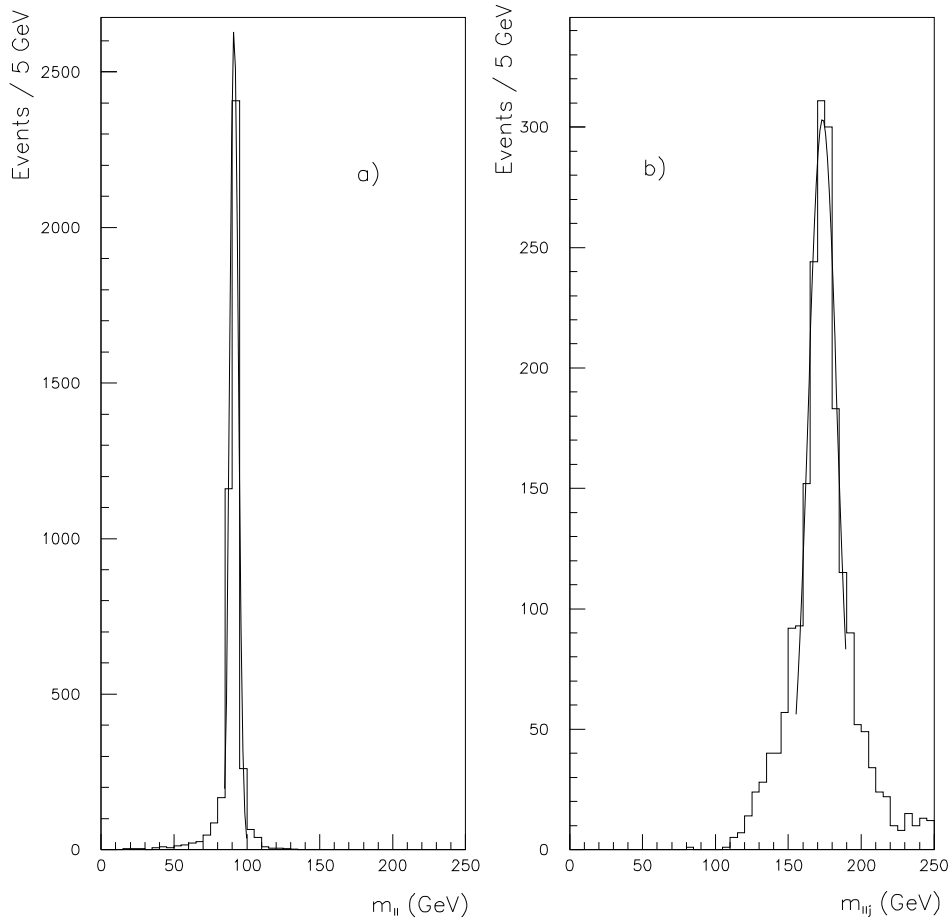,width=14cm}
\end{center}
\vspace{-0.5cm}
\caption
{a) Distribution of reconstructed invariant mass of the lepton pairs,
$m_{ll}$ for the best combination (hadronic mode).
b) Distribution of reconstructed invariant mass of
$t \rightarrow llj$ for the best combination of $llj$ (hadronic mode).}
\end{figure}

\begin{figure}[hbpt]
\begin{center}
\epsfig{file=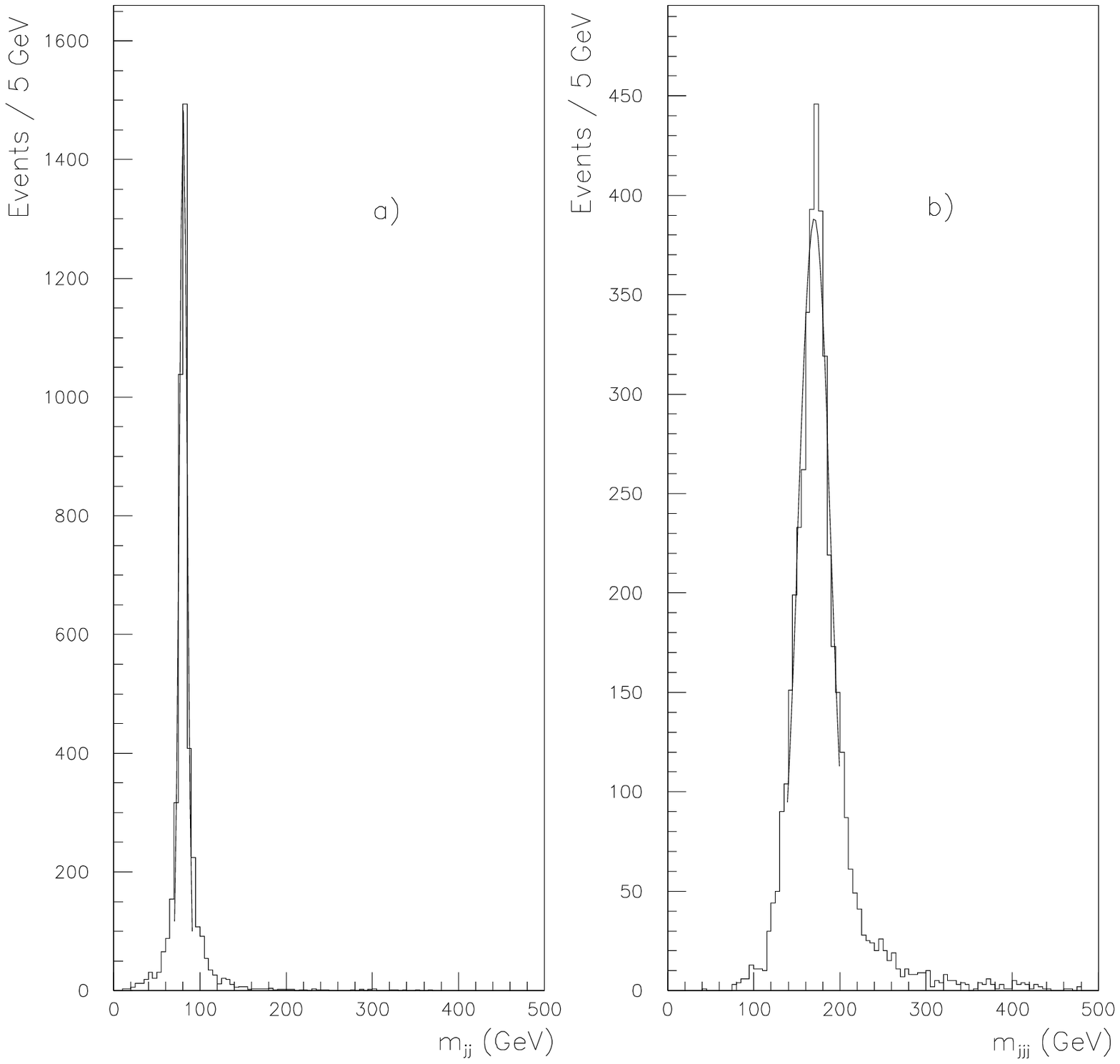,width=14cm}
\end{center}
\vspace{-0.5cm}
\caption
{a) Distribution of reconstructed invariant mass of the jet pairs,
$m_{jj}$ for the best combination (hadronic mode).
b) Distribution of reconstructed invariant mass of
$t \rightarrow jjb$ for the best combination of $jjb$ (hadronic mode).}
\end{figure}

\begin{table}[hbpt]
\begin {tabular}{|l||c|c||c|c||c|c|}\hline
\multicolumn{1}{|c||}{Description}&\multicolumn{2}{|c||}
{$t \rightarrow Zc$}&\multicolumn{4}{|c|}{Background Processes}
\\ \cline{4-7}
\multicolumn{1}{|c||}{of}&\multicolumn{2}{|c||}
{Signal}&\multicolumn{2}{|c||}{Z+jets}&
\multicolumn{2}{|c||}{Z+W} \\ \cline{2-7}
\multicolumn{1}{|c||}{Cuts} & Nevt & Eff (\%) & Nevt & Eff (\%) 
& Nevt & Eff (\%) \\ \hline \hline
Nevt gen. &19002& &2.1$\cdot 10^{7}$& &1.2$\cdot 10^{5}$ & \\ \hline
Preselection & 8742 & 46.0 & 7.5$\cdot 10^{5}$ & 3.5 & 4970 & 4.1 \\ \hline
2 leptons & 7174 & 37.7 & 5.9$\cdot10^{5}$ & 2.8 & 4456 & 3.7 \\ \hline 
4 jets & 2896 & 15.2 & 63478 & 0.29 & 400 & 0.3 \\ \hline 
${\Delta}R_{jj} > 0.4 $ & 2828 & 14.9 & 60421 & 0.28 & 390 & 0.3 \\ \hline 
${\Delta}R_{lj} > 0.4 $ & 2826 & 14.9 & 60394 & 0.28 & 361 & 0.3 \\ \hline
$M_{Z} \pm 6~GeV$ & 2426 & 12.8 & 50973 & 0.24 & 268 & 0.2 \\ \hline
$M_{W} \pm 16~GeV$ & 1006 & 5.3 & 14170 & 6.6$\cdot10^{-2}$ 
& 139 & 0.1 \\ \hline
one $b$-tag & 432 & 2.2 & 1379 & 6.4$\cdot10^{-3}$ 
& 11 & 9.1$\cdot10^{-3}$ \\ \hline
$M_{Wb} = M_t \pm 8~GeV$& 106 & 0.6 & 90 & 4.2$\cdot10^{-4}$ 
& 1 & 8.3$\cdot10^{-4}$ \\ \hline \hline
$M_{Zq} = M_t \pm 8~GeV$& 46 & 0.2 & 0 & 0
& 0 & 0 \\ \hline
$M_{Zq} = M_t \pm 12~GeV$& 58 & 0.3 & 1 & 4.8$\cdot10^{-6}$ 
& 0 & 0 \\ \hline
$M_{Zq} = M_t \pm 24~GeV$& 74 & 0.4 & 2 & 9.3$\cdot10^{-6}$ 
& 0 & 0 \\ \hline
\end{tabular}
\caption{The numbers of events and efficiencies ($\%$) of selection cuts
applied in sequence for the signal in the hadronic decay mode.}
\label{tab:hadronic}
\end{table}

 In Fig. 7.b the distribution of reconstructed  t$\rightarrow$Zq invariant
top mass $m_{llj}$ for the best combinations of $llj$
is presented for the signal. The resolution $\sigma$ of $m_{llj}$ distribution
is $\sigma_{m_{llj}}$ = 9.9 GeV.
For the reconstruction of the top mass for the decay
t$\rightarrow$Zq, the same mass windows have been chosen as for the leptonic
mode: $m_{Zq}$ $\pm 8$ GeV (narrow cut), $m_{Zq}$ $\pm 12$ GeV ($\sim ~ \sigma$)
$m_{Zq}$ $\pm 24$ GeV ($\sim ~2\sigma$). 

It is worth mentioning that, in the mass window
$m_{Zq}$ $\pm 8$ GeV, 91$\%$ of the accepted events contain $c$-jets,
t$\rightarrow llc$ and only 10$\%$ of events are reconstructed with
light jets. By widening the mass window,
the mixture of events reconstructed with light jets increases from
10\% to 22\%, but the number of events reconstructed with $c$-jets
increases too.

The analysis cuts reduce the $WZ$ background to a negligible level
in three $Zj$ mass windows. The $Z+jets$ background
vanishes in mass window $m_{Zq}$ $\pm 8$ GeV, while 
one event is accepted in $m_{Zq}$ $\pm 12$ GeV,  and two events in
$m_{Zq}$ $\pm 24$ GeV. 

The results in Table \ref{tab:hadronic} imply that
a value of Br(t $\rightarrow$ Zq ) as low as  5.9x10$^{-4}$
could be discovered at the 5$\sigma$ level with an integrated luminosity
100 fb$^{-1}$.

\section{Conclusions}

We have studied the ATLAS sensitivity to the FCNC top quark rare decay
$t \rightarrow Zq$ ($q=u,c$) for an integrated luminosity of $100~fb^{-1}$.  

\begin{table}[hbpt]
\begin {tabular}{|c||c|c||c|} \hline
\multicolumn{1}{|c||}{Window}&\multicolumn{3}{|c|}{Sensitivity to
Br(t$\rightarrow$Zq)}\\ \cline{2-4}
\multicolumn{1}{|c||}{for $M_{Zq}$}&\multicolumn{1}{|c|}{Leptonic Mode}
&\multicolumn{1}{|c||}{Hadronic Mode}&\multicolumn{1}{|c|}{Combined Result}
\\ \hline \hline
$M_t \pm 8$ GeV &  2.80$\cdot 10^{-4}$ & 
 9.60$\cdot 10^{-4}$ 
&  2.60$\cdot 10^{-4}$ \\ \hline 
$M_t \pm 12$ GeV &  2.34$\cdot 10^{-4}$ & 
 6.70$\cdot 10^{-4}$ 
&  2.18$\cdot 10^{-4}$ \\ \hline 
$M_t \pm 24$ GeV &  2.01$\cdot 10^{-4}$ & 
 5.86$\cdot 10^{-4}$ 
&  1.80$\cdot 10^{-4}$ \\ \hline 
\end{tabular}
\caption{Summary of the results.}
\label{tab:summary}
\end{table}

As summarized in Table \ref{tab:summary},
the results demonstrate that, in the leptonic mode, a branching ratio
as low as 2.0x10$^{-4}$ and, in the hadronic mode,
as low as 5.9x10$^{-4}$ could be discovered at the 5$\sigma$ level 
with an integrated luminosity of 100 fb$^{-1}$. 

Table \ref{tab:compare} compares the relative efficiences
of some of the important kinematic cuts for leptonic and hadronic 
modes of the signal. One can see that the
relative efficiences of leptons, jets, $Z$ mass, $b$-jet  and
top mass cuts for leptonic and hadronic modes are in good
agreement, as expected.  This agreement supports the consistency of
the two analyses, despite different versions of the ATLFAST code
being used in the two cases.

\begin{table}[hbpt]
\begin {tabular}{|l|c|c|}\hline
\multicolumn{1}{|c|}{ C U T S}&\multicolumn{1}{|c|}{LEPTONIC MODE}&
\multicolumn{1}{|c|}{HADRONIC MODE}\\ \cline{2-3}
\multicolumn{1}{|c|}{}& \multicolumn{1}{|c|}{Rel.Effic. ($\%$)}&
\multicolumn{1}{|c|}{Rel.Effic. ($\%$)}\\ \hline \hline
3l with $P_{T}^{l} > 20~ GeV$ in $|\eta^{l}| < 2.5 $ &70.1&  \\ \hline
2l with $P_{T}^{l} > 20~ GeV$ in $|\eta^{l}| < 2.5 $ &&82.0 \\ \hline
2 isolated jets with $P_{T}^{jet} > 50~ GeV$ &48.6&   \\ \hline
4 isolated jets with $P_{T}^{jet} > 50~ GeV$ & &39.4 \\ \hline
Lepton-jets isolation: ${\Delta}R_{lj} > 0.4 $ &94.0&99.0\\ \hline \hline
Z mass $M_{Z} \pm 6~ GeV$ &84.9&85.8\\ \hline
one tagged $b$ jet in the event &48.6&42.9\\ \hline
t$\rightarrow$Zq mass : $M_{Zq} \pm 8~ GeV$&43.4&43.4\\ \hline
t$\rightarrow$Zq mass : $M_{Zq} \pm 12~ GeV$&49.2&54.7\\ \hline
t$\rightarrow$Zq mass : $M_{Zq} \pm 24~ GeV$&75.3&69.8\\ \hline
\end{tabular}
\caption{The relative efficiencies ($\%$)
of some of the kinematic cuts for leptonic and hadronic modes of the signal.}
\label{tab:compare}
\end{table}

Combining the results from the two decay modes extends the 5$\sigma$ discovery 
potential down to a branching ratio as low as 1.8x10$^{-4}$.

\section*{Acknowledgements}

We are very indebted to D.Froudevaux, J.Parsons, M.Cobal,
E.Richter-Wass, S.Slabo-\\spitsky for very interesting and important 
discussions. We are very grateful to R.Mehdiyev for valuable advice.
We would like to thank P.Jenni and T.Grigalashvili for their  continuous
support and encouragement during our work, and J.Khubua for providing
the opportunity to proceed with this work in the future.
The authors wish to thank Z. Menteshashvili for help during
the preparation of the article.


\begin{thebibliography}{99}
\bibitem{}{F.Abe et al. (CDF Collaboration), Phys.Rev.Lett. 74 (1995) 2626;
S.Abachi et al. (DO Collaboration), Phys.Rev.Lett. 74 (1995) 2632}
\bibitem{}{D.Carlson, E.Malkawi, and C.Yuan, Phys.Lett B337 (1994) 145; and
references therein}
\bibitem{}{D.Atwood, A.Kagan,and T.Rizzo, SLAC-PUB-6580 (hep-ph/9407408) and
references therein}
\bibitem{}{H.Fritzsh, Phys.Lett. B224 (1989) 423}
\bibitem{}{T.Han, R.Peccei and X.Zhang, Nucl.Phys. B454 (1995) 527}
\bibitem{}{B.Grzadkowski, J.Gunion and P.Krawczyk, Phys.Lett. B268 
(1991) 106;\\
M.Luke and M.Savage, Phys.Lett. B307 (1993) 387;\\
G.Couture, C.Hamzaoui and M. Konig, Phys. Rev. D52 (1995) 1713}
\bibitem{}{G.Eilam, J.Hawett and A.Soni, Phys.Rev. D44 (1991) 1473}
\bibitem{}{C.S.Li, R.Oakes and J.M.Yang, PHys.ReV. D49 (1994) 293;\\
G.de Divitiis, R.Petronzio and L.Silvestrini, Nucl.Phys. B504 (1997) 45:\\
G.Couture, M.Frank and H.K$\ddot{o}$nig, Phys.Rev. D56 (1997) 4213}
\bibitem{}{F.Abe et al. (CDF Collaboration), Phys.Rev.Lett. 80 (1998) 2525}
\bibitem{}{ATLAS Collaboration, Technical Proposal, CERN/LHCC/94-43, 1994}
\bibitem{}{E.Richter-Was,
D.Froidevaux and L.Poggioli, ATLAS Internal Note ATL-PHYS-98-131, 1998}
\bibitem{}{R.Ronciani et al., Nucl.Phys. B529 (1998) 424}
\bibitem{}{J.Dodd, S.McGrath,  and J.Parsons, ATLAS Internal
           Note ATL-COM-PHYS-99-039 (1999) }
\end{thebibliography}
\end{document}